\newif\ifpreprint
\preprintfalse
\ifpreprint
   \documentclass[prl,preprint,showpacs,preprintnumbers,endfloats,amsmath,amssymb]{revtex4}
\else
   \documentclass[prl,twocolumn,noshowpacs,preprintnumbers,amsmath,amssymb]{revtex4}
\fi

\usepackage{graphicx}
\DeclareGraphicsExtensions{.eps,.ps}

\usepackage{dcolumn}
\usepackage{bm}

\begin{document}

\newlength{\figwidth}

\setlength\figwidth{82mm}


\title{Comment on ``Nonlinear current-voltage curves of gold quantum point contacts''
[Appl. Phys. Lett. 87, 103104 (2005)]}

\author{J.\ B\"urki}
\author{C.\ A.\ Stafford}%
\affiliation{%
Physics Department, University of Arizona, 1118 E.\ 4th Street, Tucson, AZ 85721
}%
\author{D.\ L.\  Stein}
\affiliation{Department of Physics and Courant Institute of Mathematical Sciences, New York University, New York, NY 10003}


\date{\today}

\ifpreprint
  \begin{abstract}
  This is a comment on the Letter on ``Nonlinear current-voltage curves of gold quantum point contacts,'' by M.\ Yoshida, Y.\ Oshima 
and K.\ Takayanagi [Appl. Phys. Lett. 87, 103104 (2005)].
  \end{abstract}
\fi

\maketitle

In a recent letter \cite{yoshida05}, Yoshida {\it et al}.\ report that
nonlinearities in current-voltage curves of gold nanowires
occur as a result of a shortening of the distance between the
electrodes at finite bias, which induces a thickening of the wire, and a
corresponding increase in conductance.
Although they find that the electrode displacement occurs for all wires, the
induced thickening of the wire, and corresponding
nonlinearity of the {\sl I-V} curve, is only observed for short contacts, while
long wires keep a constant radius, and have a linear {\sl I-V}.
Yoshida {\it et al}.\ provide a plausible quantitative explanation for the shortening of the nanowire
at finite bias in terms of thermal expansion of the electrodes induced by Joule heating,
but do not offer any explanation for 
the difference in behavior of long and short wires.
We argue here that electron-shell effects \cite{burki05}, which favor wires with
certain ``magic radii,'' 
prevent the thickening of long wires under compression, but have little effect
on wires below a critical length \cite{burki05a}. 

\ifpreprint
  \begin{figure}[p]
\else
  \begin{figure}[bt]
    \vspace{1mm}
    \includegraphics[width=\figwidth]{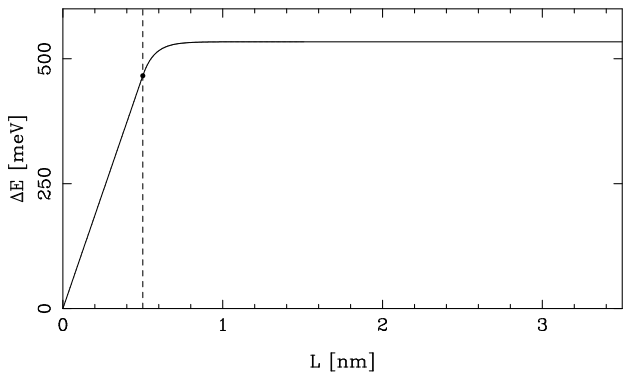}
    \vspace{-2mm}
\fi
    \caption{The activation energy $\Delta E$ as a function of wire length $L$ for 
a magic Au nanocylinder with $G=17 G_0$.  
The vertical dashed line indicates the critical length $L_c$ at which the transition in activation
behavior takes place.
   \label{fig:barrier}}
\ifpreprint\relax\else\vspace{-6mm}\fi
\end{figure}

The stability and structural dynamics of simple metal nanowires have been systematically
investigated in a continuum model known as the {\em nanoscale free-electron model}
\cite{burki05,burki05a,burki03}.
This model includes electron-shell effects, which are crucial for the stability of long, thin
wires \cite{burki05}, and permits the study of long-timescale collective phenomena 
\cite{burki03,burki05a}, which would be inaccessible to atomistic simulations.
Cylindrical metal nanowires with certain magic radii are predicted to be metastable
structures \cite{burki05,burki05a};
their lifetimes, i.e.\ the time it takes a wire to spontaneously jump to the
next magic radius under thermal fluctuations, have been computed \cite{burki05a}. 
The escape process is found to undergo a transition as a function of wire
length, with short wires having a homogenous transition 
state and an escape barrier proportional to wire length, while longer
wires escape through nucleation of an instanton that propagates along the wire,
with an escape barrier independent of wire length (see Fig.\ \ref{fig:barrier}).
The critical length at which this transition occurs is found to be $L_c \sim
0.5$ nm for a gold wire with a conductance $G = 17\,G_0$ 
($G_0=2e^2/h$ is the conductance quantum).

\ifpreprint
  \begin{figure}[p]
\else
  \begin{figure}[tb]
   \vspace{1mm}
   \includegraphics[width=\figwidth]{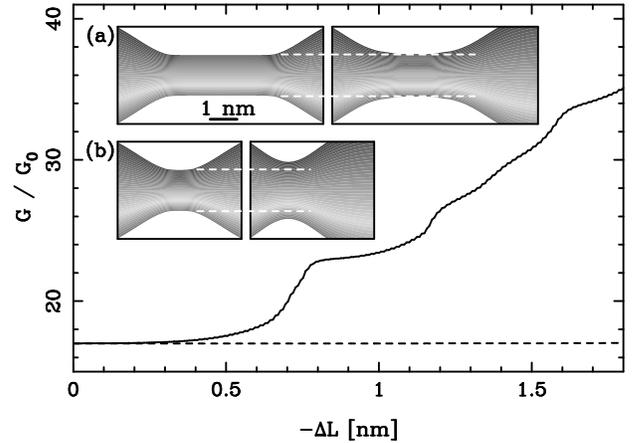}
   \vspace{-2mm}
\fi
   \caption{Conductance $G$ versus length change $\Delta L$, during compression of short ($L\sim1$ nm, solid line) and long ($L\sim 4$ nm, dashed line) Au wires, both with initial conductance $G = 17\,G_0$.
   Inset (a) shows the long wire before (left) and after (right) expansion of the leads; 
   the corresponding shapes for the short wire are shown in inset (b). 
   \label{fig:wires}}
\ifpreprint\relax\else\vspace{-6mm}\fi
\end{figure}

We believe that this transition is responsible for the observed difference in
behavior of long and short wires in Ref.\ \onlinecite{yoshida05}: The length of
the long wire ($\sim 5$ nm) is much larger than the critical length $L_c$, and
there is therefore a large energy barrier preventing a change of radius.
On the other hand, the short wire ($L \sim 1$ nm) is only slightly longer than
$L_c$, so that the electrodes' displacement shortens it near $L_c$, where the
energy barrier decreases rapidly and the wire becomes `soft' to radius fluctuations. 
Moreover, below $L_c$ the transition state is uniform, and can be accessed
directly by strain, through elastic deformation, while for wires longer than
$L_c$, the nonuniform transition state is accessible only via rare large thermal
fluctuations.

This picture is confirmed by simulations of the structural dynamics of a
nanowire under compression (see  Fig.\ \ref{fig:wires}).
The nanostructure, approximated by a continuous medium, is assumed to evolve
under surface self-diffusion of metal atoms, and electron-shell effects are
included in the energy functional (see Ref.\ \cite{burki03} for details of the
model).
The electrode displacement $\Delta L$ causes a short wire to thicken, leading to
an increase of its electrical conductance (solid line),
whereas a long wire is only shortened without any change of its radius or
electrical conductance (dashed line); 
its radius is essentially ``pinned'' at a deep minimum of the electron-shell potential \cite{burki05}.
The calculated increase of $dI/dV$ for the short wire in Fig.\ \ref{fig:wires} is quantitatively 
consistent with the observations of Ref.\ \onlinecite{yoshida05}.

Note that the qualitatively different behavior of short and long wires would not occur if they acted as solid rods: 
Under a purely elastic deformation, a 5nm-long Au rod would increase in radius by 10\% when shortened by
1nm, leading to a 35\% increase in its (quantized) electrical conductance.
This provides further evidence that nanowires, 
whose energetics are dominated by surface- and electronic quantum-size effects \cite{burki05}, do not behave as classical elastic solids.

This work was supported by NSF Grant Nos. 0312028 and 0351964.
\ifpreprint\relax
\else\vspace{-4mm}\fi

\ifpreprint\newpage\fi

\ifpreprint
  \newpage
  \printfigures
\fi

\end{document}